\documentclass[showpacs,preprintnumbers,amsmath,amssymb]{revtex4}
\usepackage{graphicx}

\begin{document}
\title{Nonlocal competition and logistic growth: patterns, defects and fronts}
\author{Yosef E. Maruvka and Nadav M. Shnerb}
\affiliation{Department of Physics, Bar-Ilan University, Ramat-Gan
52900 Israel}
\begin{abstract}
Logistic growth of diffusing reactants  on spatial domains with
long range competition is studied. The bifurcations cascade
involved in the transition from the homogenous state to a
spatially modulated stable solution is presented, and a
distinction is made between a modulated phase, dominated by single
or few wavenumbers, and the spiky phase, where localized colonies
are separated by depleted region. The characteristic defects in
the periodic structure are presented for each  phase, together
with the invasion dynamics in case of local initiation. It is
shown that the basic length scale that controls the bifurcation is
the width of the Fisher front, and  that the total population
grows as this width decreases. A mix of analytic results and
extensive numerical simulations yields a comprehensive examination
of the possible phases for logistic growth in the presence of
nonlocal competition.
\end{abstract}


 \maketitle

\section{Introduction}

\noindent Recently, there is a growing  interest in the spatial
properties of logistic growth  with nonlocal interactions
\cite{Sakai, Doebeli, Bolker, Tokita, Hoopes, Anderson, Sayama,
Fuentes, Birch, shnerb, Garcia}. A variety of models have been
introduced, including various types of interaction kernels,
deterministic and stochastic evolution and growth or death rate
that depends on the local population. A common feature found in
all these models is the \emph{segregation transition}, i.e., for
small enough diffusion and for certain interaction kernels the
homogenous state of the system becomes unstable and the steady
state is spatially heterogenous. This feature turns out to be
stable against the stochasticity induced by the discrete nature of
the reactants, and the total carrying capacity (per unit volume)
of the stochastic system depends on the details of the spatial
segregation \cite{Birch, Garcia}.

In previous work \cite{shnerb}, the general conditions for the
integral kernel to allow for spatial segregation have been
presented, and the existence of topological defects between
ordered domains has been analyzed in detail for a logistic growth
on a one dimensional array of patches with nearest-neighbor
competition. Here, a comprehensive study of  this
reaction-diffusion equation  is presented: short-range
interactions are shown to yield spatial modulation of arbitrary
large wavelength and different type of defects, the total
population of the system admits nontrivial dependence upon the
diffusion rate, and the dynamics of the system is studied, both
for global initiation and for local initiation. The appearance of
domains with different order parameter and the features of the
boundaries between them is considered in detail for various
situations.

Our starting point is the well-investigated Fisher-KPP equation
\cite{fisher, kol}, first introduced by Fisher to describe the
spread of a favored gene in population:
\begin{equation}
\frac{\partial c(x,t)}{\partial t}= D \nabla^2
c(x,t)+ac(x,t)-bc^2(x,t).
\end{equation}
Clearly, this equation is a straightforward generalization of the
logistic growth to spatial domains, and allows for two steady
states: an unstable state with  $c(x)=0 \ \forall x$ and the
stable steady state $c(x)=a/b$. It was shown that, for any local
initiation of the instability (i.e., $c(x)\neq 0 $ on a compact
domain) the invasion of the stable phase into the unstable region
takes place via a front that moves in a constant velocity $v_F = 2
\sqrt{Da}$. The stability of this solution, the fact that the
velocity is determined by the leading edge ("pulled front")  and
the corrections to this expression due to stochastic noise
associated with the discrete nature of the reactants
\cite{derrida} has been reviewed, recently, by various authors
\cite{saarloos}.

The FKPP equation is the simplest equation that describes the
transition from unstable to stable steady state  on spatial
domains, and as such it fits many situations, from the spread of a
disease by infection to the advance of a fire or new technology.
Accordingly, this model has been widely studied from many points
of view and has been generalized in many directions such as
modified interaction terms, non linear diffusion and so on.

The process considered here, logistic growth with nonlocal
competition,  is described by the generalized FKPP equation:
\begin{equation} \label{eq2}
\frac{\partial c(x,t)}{\partial t}= D \nabla^2
c(x,t)+ac(x,t)-c(x,t)\int^\infty_{- \infty} \gamma(x,y) c(y,t)dy,
\end{equation}
where $\gamma(x,y)$ is the interaction kernel, and the original
FKPP process corresponds to the limit $\gamma(x,y)=\delta(x-y)$.

The  motivation for the study of this process comes from one of
the basic mechanisms in population growth, namely, the competition
for common resource. In any autocatalytic system the
multiplication of agents depends on various resources (energy,
chemicals, water etc.). If there is only limited amount of the
resource, its consumption leads to extinction, so generally any
crucial resource should be  deposited, and its availability
dictates the saturation value for the population. As a concrete
example let us look at vegetation \cite{merron, lavee}: the common
resource needed for vegetation is water, and the rain corresponds
to deposition of this resource. If the resource dynamics is much
faster than that of the  agents (shrubs, trees etc.), there is, at
any time, a soil moisture profile that reflects the instantaneous
vegetation configuration, and there is a depletion of this
moisture at the spatial region around a biomass unit. Accordingly,
the environmental conditions for a new agent at this region
becomes hostile. Following arguments of this type one suggests
that \emph{competition for common resource induces  long range
competition among agents via the depletion of the resource in the
surroundings of an agent}. Another examples may involve the
competition for light \cite{huisman} and cooperation among agents
(symbiosis)  that may yield "negative competition" among the
reactants.

Numerical simulations of the  dynamics corresponds to Eq.
(\ref{eq2}) require space and time discretization. In this work
the time evolution of the system is generated via forward Euler
integration, where the time step is taken small enough such that
further reduction of it do not effect the results. We simulate a
system of discrete patches, where the hopping rate is proportional
to the diffusion constant.

Let us present some a-priori considerations related to this
system. There are few basic types of steady state solutions:
first, it may happens that the steady state is \emph{homogenous}:
this may be the case if the long range competition is too small,
or if the interaction kernel do not allow for the instability to
occur \cite{shnerb, Sayama}. At some point in the parameter region
a bifurcation may occur, and the homogenous states becomes
unstable to modulation of wavelength $2 \pi /k$. Right above the
bifurcation one expects, though, to see an inhomogeneous
(modulated) steady state. Far from the bifurcation point there are
many unstable wavelengthes  and some sort of mode competition
takes place. In the other limit, i.e., very strong competition,
one may expect that "life" at a single patch forces all the other
patches at finite range to be (almost) empty, so the steady state
is sort of "spiky" phase, where many active wavelengthes
participate in the formation of localized bumps.

As we are looking at a dynamic system with no noise, few stable
steady states may exist simultaneously, each admits its own basin
of attraction in the space of possible initial conditions.
Numerically, however, it turns out  that only one important
distinction should be made, namely, between local and global
initiation: the initiation is "local" if at $t=0$ there is finite
support to the colony, while if the system begins with random
small biomass that spreads all around it corresponds to global
initiation.  Within each of these subclasses, the numerics
suggests that a generic initial condition will flow into a unique
steady state.

This paper is organized as follows: in the second section the
stable spatial configurations (steady states stable against small
fluctuations) are presented: the conditions for an instability of
the homogenous solution are reviewed and discussed, and the
properties of the final state are identified in different
parameter regions, leading to a characteristic "phase diagram". In
the next section the appearance of defects (separating spatial
regions with different order parameter phase) is studied. The
fourth section deals with the "spiky" phase, where many excited
modes superimposed to yield a pattern of spikes and the typical
defect is a combination of two depletion regions.  In the fifth
section there is a brief description of phases and defects in two
spatial dimensions, and next  the effect of the spatial
segregation on the global population is considered. In the seventh
section the dynamic properties of the model are discussed,
including the velocity of the primary and the secondary Fisher
fronts and the appearance of topological defect in the invaded
region. Some comments and conclusions are presented at the end.

\section{Static properties}

In this section we consider the  steady state solutions for
equation (\ref{eq2}) on spatial domain of coupled, identical
patches. The initiation is assumed to be \emph{global}, i.e., the
initial conditions are small, randomly spread, reactant population
at each spatial patch. The model considered here allows for
nontrivial spatial organization even in the absence of diffusion,
due to the long range competition, and global initiation helps to
see these features within reasonable simulation times. The
differences, if any, between global and local initiation will be
considered in the last section.

\subsection{The bifurcation cascade}

Let us consider the spatially discretized version of (\ref{eq2}),
i.e., an infinite one dimensional array of identical patches
coupled to each other by diffusion and  long range competition.
The time evolution of the reactant density at the n-th site,
$\widetilde{c}_n$, is given by:

\begin{eqnarray}
\label{eq:1} \frac{\partial \widetilde{c}_n(t)}{\partial
t}&=&\frac{\widetilde{D}}{l_0^2}[-2\widetilde{c}_n(t)+\widetilde{c}_{n+1}(t)+\widetilde{c}_{n-1}(t)]+
a \widetilde{c}_n(t)  \nonumber \\  &-&b \widetilde{c}_n^{2}(t) -
\widetilde{c}_n(t) \sum_{r=1}^\infty \widetilde{\gamma}_r
[\widetilde{ c}_{n+r}(t)+\widetilde{c}_{n-r}(t)].
\end{eqnarray}
where $\widetilde{D}$ is the diffusion constant  and
$a,b,\widetilde{\gamma}$ are the corresponding reaction
coefficients. One may precede to  define the  dimensionless
quantities,
\begin{equation}  \tau=at, \qquad
 c=b\widetilde{c}/a, \qquad  \gamma_r=\widetilde{\gamma}_r/b,
 \qquad
  D=\frac{\widetilde{D}}{a l_0^2}.
  \end{equation}
Note that the new   "diffusion constant"  is $D = W^2/l_0^2$,
where $W \equiv \sqrt{D/a}$ is the width of the Fisher front, so
the dimensionless diffusion is determined  by the ratio between
the front width and the lattice constant. The continuum limit,
though, is the limit where the front width is large in units of
lattice spacing. With these definitions Eq. (\ref{eq:1}) takes its
 dimensionless form,
\begin{eqnarray}
\label{eq:5} \frac{\partial c_n}{\partial \tau}&=& D
[-2c_n+c_{n+1}+c_{n-1}] \nonumber \\ &+& c_n \left( 1-c_n-
\sum_{r=1}^\infty \gamma_r [c_{n+r}+ c_{n-r}] \right),
\end{eqnarray}
that may be expressed in Fourier space  [with $A_k \equiv \sum_n
c_n e^{iknl_0}$] as,
\begin{equation} \label{eq3}
\dot{A}_k = \alpha_k A_k - \sum_q \beta_{k-q} A_q A_{k-q},
\end{equation}
where
\begin{eqnarray}
\alpha_k \equiv 1-2D[1-cos(kl_0)] \\
\beta_k \equiv 1+2\sum_{r=1}^\infty \gamma_r cos(rkl_0).
\end{eqnarray}

\noindent Following \cite{ns}, one observes that   $c_n$ is
positive semi-definite so $A_0$ is always "macroscopic". Any mode
is suppressed by $A_0$; accordingly,  for small $\gamma_r$ one
expects only the zero mode to survive. If, on the other hand,
$\gamma_r$ increases above some threshold, bifurcation may occur
with  the activization of some other $k$ mode(s), and the
homogenous solution becomes unstable. The condition for occurrence
of such bifurcation is that:
\begin{equation}\label{bif}
g(k)=\beta_k + 2 \beta_0 D [1-cos(kl_0)]< 0
\end{equation}
fulfilled by some k. This is the situation where patterns appear
and translational symmetry breaks. Right above the bifurcation
there is only one active $k$ mode that dictates the modulation of
the system. As $g(k)$ decreases further there are many active
modes that compete with each other via the nonlinear terms of
(\ref{eq3}), and the linear stability analysis of the homogenous
state may be irrelevant to the final spatial configuration.

\subsection{Nearest neighbors interactions}

In previous work, the properties of the system have been
considered for the extreme case where the competition takes place
only between neighboring sites ($\gamma_r = \gamma$ for $r=1$ AND
$\gamma_r=0$ if $r>1$). For nearest neighbors (n.n) interaction of
that type  the only stable wave number  is $k=\pi /l_0$, where
$l_0$ is the lattice constant, and the bifurcation takes place at
$\gamma<1/2$.   The spatial state at this wave number is
$u_n=A_0+A_\pi cos(n \pi / l_0)$ and the spatial structure is of
the form ...ududud... (u=up, large amount of biomass, d=down,
small amount). In the absence of diffusion spatial segregation
takes the form 101010…, i.e., only the even (odd) sites are
populated. Obviously, starting from generic random state different
domains are crated with odd or even "order parameter" and  kinks
(domain walls) emerge between different domains. As shown in
\cite{shnerb}, the structure of these topological defects,
including their size (that diverges at the segregation transition)
and their exact form may be calculated analytically.

\subsection{Next nearest neighbors (n.n.n)}

\noindent Quiet surprisingly, the increase of the competition
radius by a single site  takes us to a completely different
regime. While in the case of nearest neighbors interaction the
spatial modulation length and the competition length are the same,
next nearest neighbors competition (and, accordingly, any
interaction of longer range) may yield, upon tuning the
parameters, spatial modulation of arbitrary large wavelength. This
situation resembles the case of magnetic systems, e.g., an  Ising
chain: if the exchange interaction is only between nearest
neighbors the equilibrium state admits only an up-down
modulations, while n.n.n. interaction may yield large solitons, as
shown by \cite{bak}. In that sense the next n.n. case demonstrate
the essential features of the long range competition model in a
generic way, while at least part of the results may be inferred
analytically.

The most general form of next nearest neighbors interactions is
given by Eq. (\ref{eq:5}) with:

\begin{equation}
\gamma_r = \left\{  \begin{array}{cc}
  \gamma_1 & r = 1 \\
  \gamma_2 & r = 2 \\
  0 & \texttt{else}  \\
\end{array} \right.
\end{equation}

\begin{figure}
\includegraphics[width=7cm]{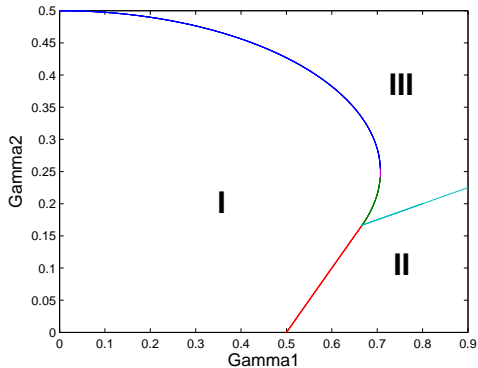}\\
\caption{Phase diagram for next nearest  neighbors competition at
$D=0$.  Region I is the homogenous, while II marks the up-down
stable solution region similar to the n.n. case. In  region  III
the  wavevector $k_3$ (defined in the text) is stable and various
wavenumbers may be active.
 }\label{fig3}
\end{figure}

The bifurcation threshold is defined now by the equation:

\begin{equation}\label{wave}
g(k)=1+2\gamma_1 cos(kl_0)+\gamma_2 cos(2kl_0)\beta_k + 2 \beta_0
D [1-cos(kl_0)]= 0.
\end{equation}

\noindent where  $g$ has extremum points at $k_{1,2}=0, \pi $ and
\begin{equation} \label{eqk}
k_3=arcos(\frac{-\gamma_1+\beta_0D}{4\gamma_2}).
\end{equation}
If a real wavevector  $k_3$ exists (i.e., at
$|\frac{-\gamma_1+\beta_0D}{4\gamma_2}|<1$) it will be the minimum
of $g(k)$ while $k_{1,2}$ are maxima.  For the range of parameters
where $k_3$ is imaginary the minima may be at $k= \pi$ and the
modulation is of "up-down" type, or  $k=0$, where the homogenous
state is stable. The resulting phase diagram, in the
$\gamma_1-\gamma_2$ plane with zero diffusion, is presented in
figure (\ref{fig3}): In region I the homogenous state is stable,
while in region II the bifurcation takes the system to the up-down
mode, like the situation for n.n. interaction. In region III,
however, $k_3$ dominate and modulations of any size may occur. The
bifurcation line is given (in the presence of diffusion) by the
two branches of the equation:

\begin{equation}
\gamma_2=\frac{1+2D-D^2+\gamma_1(5D-2D^2) \pm
\sqrt{1+4D+14D\gamma_1-2\gamma_1^2+12D\gamma_1^2}}{2D^2+4-16D}
\end{equation}
that reduces, at the $D=0$ case, to the simple form:
\begin{equation}
\gamma_2=\frac{1 \pm \sqrt{1-2\gamma_1^2}}{4}.
\end{equation}

\subsubsection{wavelength selection, mode competition and the spiky phase}

\begin{figure}
  \includegraphics[width=7cm]{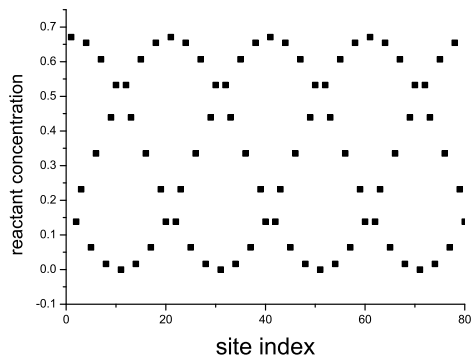}\\
  \caption{ Spatial structure of wavevector $k=
  14\pi/20 l_0$. According to Eq. (\ref{eqk})  one expects the modulation length
  to be $\lambda=20 l_0/7$, but
  but  discreteness of the
  lattice allows only for commensurate periodicity of  20 sites.
  }\label{fig4}
\end{figure}

\noindent From (\ref{eqk}) it seems that the bifurcation
wavelength is bounded from above by the interaction length. This,
however, is not the actual situation on a discrete lattice: the
wavelength inferred from Eq. (\ref{eqk}), although bounded,  is
generically incommensurate with the lattice constant, and the
system should choose a commensurate one. It turns out that, if the
wavelength is rational (i.e., if $k_3 = 2 \pi m/n$, where $m$ and
$n$ admits no common denominator) the spatial modulation repeats
itself after $n$ lattice sites. A typical example is the steady
state obtained numerically for the case $m=7,n=20$ where a
period-20 modulation appears, as demonstrated in Figure
(\ref{fig4}). At finite system the maximal $n$ allowed is of order
of the system size, and only in an infinite system all rational
fractions may be activated. Note that in an infinite system any
change of the interaction parameters yields different wavelength,
a phenomena that resembles the "devil staircase" situation in spin
systems \cite{bak}.

\begin{figure}
  \includegraphics[width=7cm]{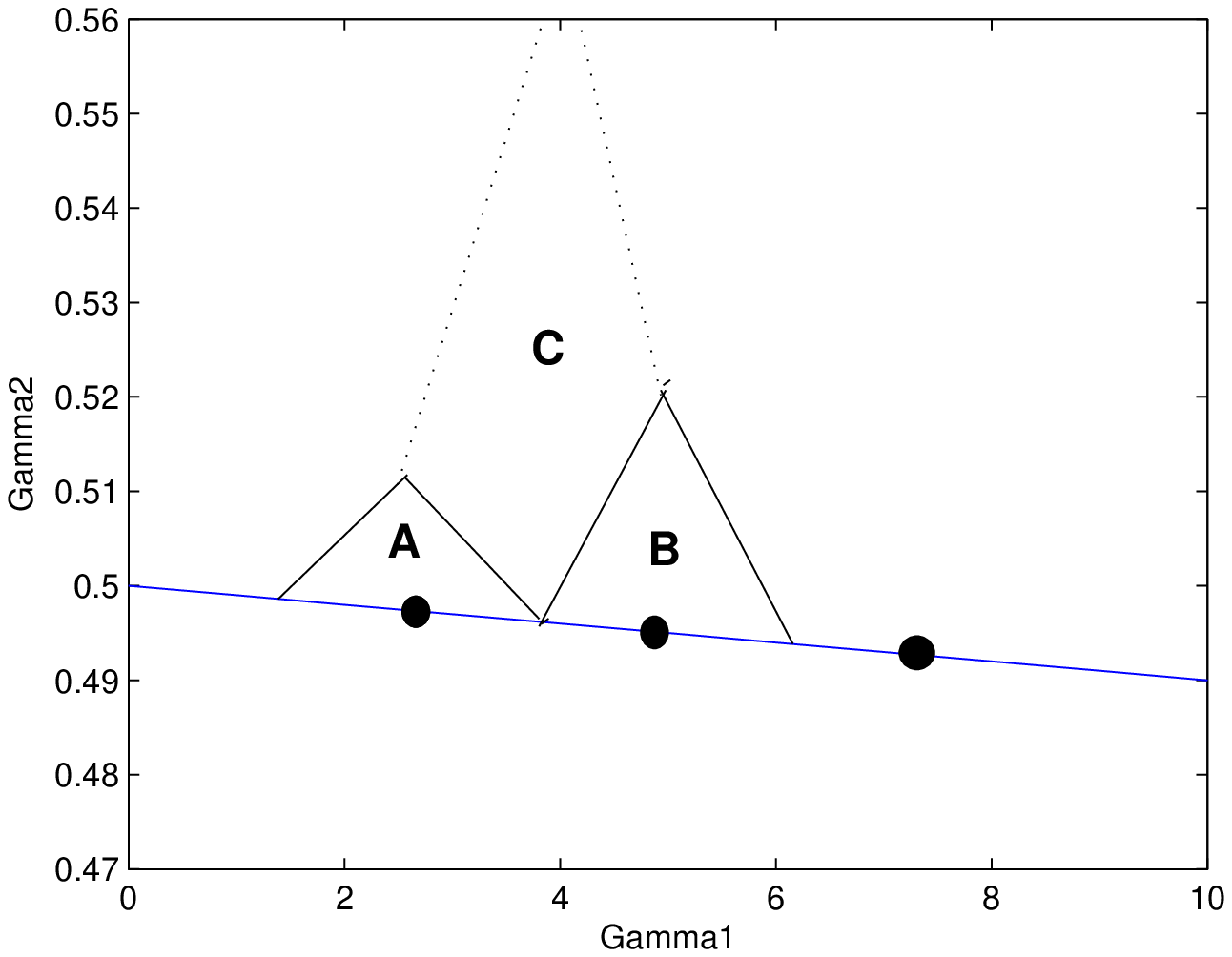}\\
\caption{Basins of attraction for allowed states close to the
bifurcation line (sketched). The straight line is an enlarged
portion of the bifurcation line (I - III interface) of figure
(\ref{fig3}). The bold points on this line correspond to an
allowed states, i.e., states with wavelength  commensurate with
the lattice size. Each possible wavevector admits a basin of
attraction, like those denoted by A and B in the figure. Starting
from $\gamma_1$,$\gamma_2$ values inside region A, for example,
the system flows to the modulation corresponds to the bold point
inside the triangle. Away from the bifurcation line (region C) few
basins of attraction overlap and mode competition takes place.
Even further away, deep in region III, the system is in the spiky
phase: many active modes exist and their superposition  yields the
"wave packet" characteristics of the  spikes}\label{fig5}
\end{figure}

For finite system, though, there is a set of points along the
bifurcation line that correspond to the allowed wavelengthes.
Numerical simulation indicates that there is a basin of attraction
around each of these points, i.e., if the interaction parameters
$\gamma_1$ and $\gamma_2$ yields a prohibited wavelength the
system flows into one of the closest allowed modulations. The
overall structure is demonstrated in figure (\ref{fig5}): close to
an isolated point there is a basin of attraction, but further away
from the bifurcation line these regions begin to overlap, and the
system flows into some mixture of the closest allowed states,
depending on its initial conditions. Deep in region III many
wavenumbers are involved; the interaction parameters are
relatively large, and instead of simple harmonic modulation the
system flows, generically, into a spiky steady where the "living
colonies" are separated by the interaction length and are not
effected by the competition between patches.

Although the numerical examples presented here are for a system
with next nearest neighbor competition and without diffusion, it
is easy to extract from it the properties of the steady state in
general. The effect of diffusion is to increase the size of the
stable region so the bifurcation line of Figure (\ref{fig3}) moves
outward together with the pure and the spiky states. For
interactions of longer range the parameter space is of higher
dimensionality but all other features are essentially  the same.

\section{Defects}

The transition from the homogenous to the modulated state involves
spontaneous breakdown of translational symmetry, and upon global
initiation one may expect domain walls, or kinks, that separate
spatial regions with different order parameter. The presence of
these defect and their character is crucial for the understanding
of the system response functions, e.g., its behavior under small
noise:  as there is no preference to one phase of the order
parameter the kinks may move freely, while the "bulk" of the
domain is much more stiff. In the following paragraphs the
characteristic defects for various phases are presented.

\subsection{Domain walls}
\begin{figure}
  \includegraphics[width=7.7cm]{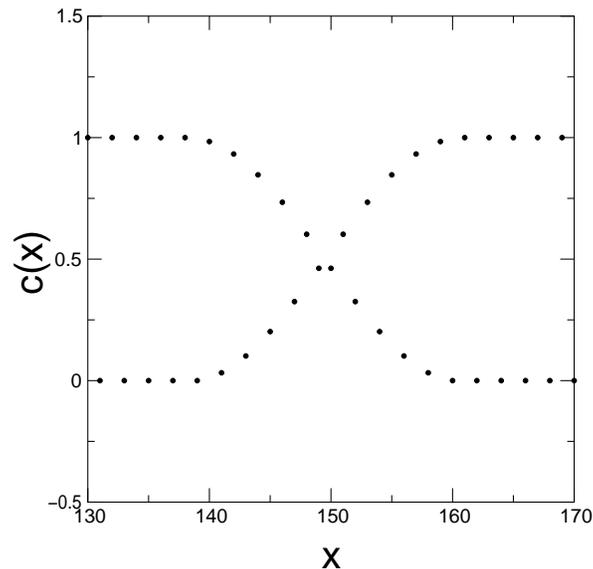}\\
  \caption{A typical kink of length $L=20$, an outcome of forward Euler integration of Eq. (\ref{eq:5})
 (with n.n. competition) on 1024 lattice points with periodic boundary conditions and random initial conditions at
   $\gamma = 0.505$ (just above the bifurcation). } \label{kink}
\end{figure}

As mentioned above, the nearest neighbors competition leads, above
the bifurcation threshold, to to appearance of an up-down
modulation ($k=\pi$), and if there is no diffusion the steady
state is the 0101010 configuration.  Clearly there are two
equivalent segregation of this type, namely, filled odd sites and
empty even sites and vise versa. Accordingly, in  case of global
initiation (random "seeds" are spread all around) one finds
domains of the stable patterns with different parity, and domain
walls (technically known as kinks or solitons) that separate these
regions, as seen in figure (\ref{kink}). The nearest neighbor
interaction is simple enough to allow for an analytic solution for
the kink, and the numerical results confirm the predictions
\cite{shnerb}.

In the presence of diffusion there is a "smearing" of the above
results: the homogenous state is stable for larger $\gamma$, and
above the segregation threshold the steady state is "smeared" from
$...01010...$ to an "up - down - up -down" form, and the kinks are
not of finite size but admit exponentially decaying tails. See
\cite{shnerb} for details.

\subsection{Phase shift}

\noindent Unlike the nearest neighbor case, competition of longer
range leads to instabilities with wavelength of more than one
site, i.e., $c_n=A_0+A_k cos(n k l_0)$ with general $k$. This
opens the problem of defects between ordered regions. Inspired by
the nearest neighbors example one may expect  another types of
kinks that separate different regions of ordered state.
Surprisingly, this is not the case. Instead of getting kinks
between different oriented regions of the activated wavenumber,
one gets \emph{single} oriented region with \emph{phase shift},
namely, the spatial structure is of the form $c_n=A_0+ B cos(n k
l_0+\phi)$, where $\phi$ is the phase shift between the actual
solution and the predicted modulation $c_n=A_0+A_k cos(n k l_0)$
and $B = A_k/cos \phi$.

\begin{figure}
  \includegraphics[width=7cm]{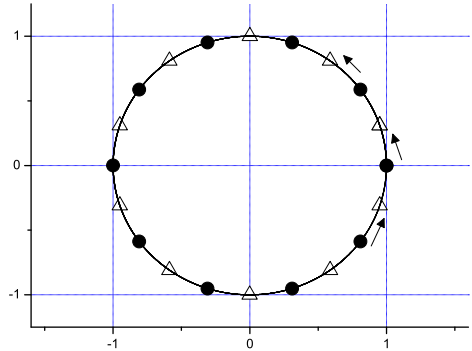}\\
 \caption{ The set of different values of population size at different site is
 presented on the unit circle where the linear analysis predicts
 an instability with wavenumber  $k=3\pi/5l_0$.
 The filled  circles are the values of $cos(3n \pi/5l_0)$, the solution predicted by the
 naive argument, and this is indeed a stable solution with finite
 basin of attraction [see Figure (\ref{pure})].
 It turns out, however, that generic initial conditions flow into a phase shifted solution where the
 population of the form  $cos(3n\pi/5l_0+\phi)$  (shown in Figure (\ref{shift}).
 The value of $\phi$ is half of the angular distance between two close
 sites, here corresponds to the open triangles on the unit cycle.
}\label{cycle}
\end{figure}

\begin{figure}
  \includegraphics[width=7cm]{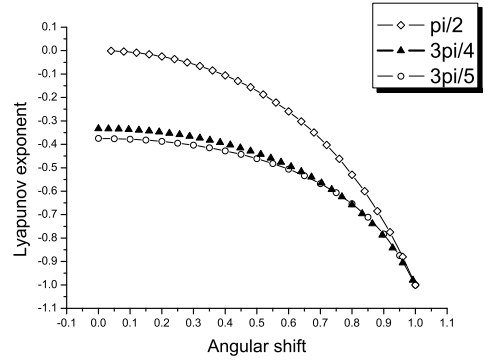}\\
  \caption{The Lyapunov exponent (in arbitrary units)  for states of the form $cos(nk+\phi)$
 is shown against  $\phi$  (in units of $k/2$) for various wavenumbers. While the steady state is
 stable for any $\phi$, the most stable state corresponds to $k/2$, and  }\label{stability}
\end{figure}

\begin{figure}
  \includegraphics[width=7cm]{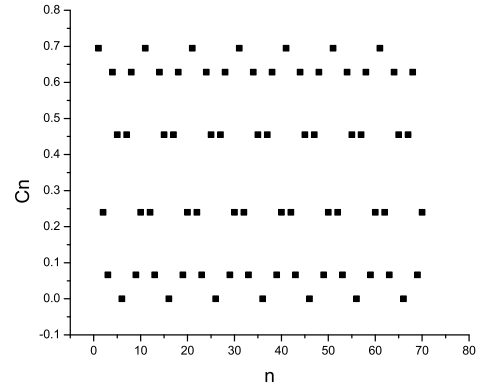}\\
  \caption{The steady state for the same parameters of figure (\ref{cycle}), where the initial
  conditions are close to the $\phi=0$ solution, i.e.,
  $c_n(t=0) = A_0+A_k cos(3\pi n/5) + \delta_n$, where $\delta_n$ is a small random number.
  The system flows into the  $\phi = 0$ case, in agreement with the local stability analysis
  presented in
 Figure (\ref{stability}).    }\label{pure}
\end{figure}

\begin{figure}
  \includegraphics[width=7cm]{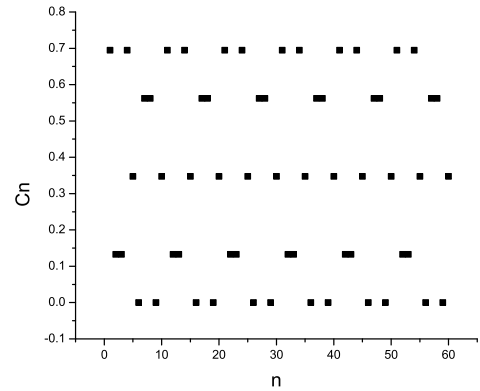}\\
  \caption{Same as Figure (\ref{pure}), but now the initial conditions are generic  $c_n(t=0)=\delta_n$.
  The system flows to the most stable steady state that corresponds, in this case, to
  $c_n=A_0+B cos(3 n \pi/5 + \phi)$, with $\phi = 3 \pi / 10$.   }\label{shift}
\end{figure}

On the unit cycle (Figure [\ref{cycle}]) the meaning of this
additional phase  is a shift of all  point by $\phi$. This shift
may reduce the number distinct values in one cycle  by one, as
indicated in the example of  Figure (\ref{cycle}): here, instead
of six distinct  values  taken by $c_n$ along one wavelength,
there are only five. Both numerical simulations of the system
dynamics, starting from random initial conditions, and stability
analyzes of the possible steady state for arbitrary $\phi$
indicates that, although any $\phi$ corresponds to locally stable
solution, the most stable $\phi$ equals to half of the angular
distance between two adjusting points on the circle. In Figure
(\ref{cycle}) the actual phase shifted pattern is shown for $k=3
\pi /5 l_0$, while Figure (\ref{stability}) indicates that the
most stable phase corresponds to $\phi/\phi_0 = 1/2$.  As the
Lyapunov exponent of any $\phi$ is negative,  small perturbations
around any $\phi$ value (in particular, $\phi =0$) decay. Figure
(\ref{pure}) shows the corresponding stable mode with $\phi=0$
where the initial conditions are small perturbation around it.
Figure (\ref{shift}), on the other hand, shows the final state
with generic initial conditions, where the system flows to the
most stable pattern with $\phi/\phi_0 = 1/2$.

\section{The spiky phase}

Deep in region III of the phase diagram [Figure (\ref{fig3})] many
wavevectors are excited, with strong mode competition between
them, and the linear analysis picture based on Fourier
decomposition becomes ineffective. Better insight into the system
comes from a real space analysis: deep into region III the long
range competition is strong, and within the effective interaction
range new colony can not develop in the presence of a fully grown
one. Accordingly, this phase is characterized by fully developed
colonies separated by "dead regions" of constant length that
reflects the effective interaction length. In Fourier space, this
corresponds to many active modes that build together a periodic
structure of "bumps".

In case of global initiation, of course, defects may appear in the
stable steady state as the system flows to different order
parameter in different regions. Again, it is better to use the
real space picture in order to describe these defects. The
situation is close to what observed in the case of random
sequential adsorption  \cite{ads,lavee}: while an "optimal"
filling of the system admits a periodic structure of living
patches with periodicity of, say, $L$ lattice points,  it may
happens that the distance between two fully developed sites is
between L and 2L, and all the site in between should remain empty
due to the long range competition. The emerging spatial
configuration is of ordered regions (with coherence size that
depends upon the dynamics) separated by "domain walls", where the
width of these walls is taken from some distribution function
between zero and the interaction effective length.

\section{Two dimensional system}

Although all the analysis presented was in one dimension, the
basic picture is the same for higher dimensionality. In
particular, the bifurcation condition is similar, nearest neighbor
interactions yields a "checkerboard" phase above the bifurcation
line,  and the  spiky phases is also observed.

\begin{figure}
  \includegraphics[width=7cm]{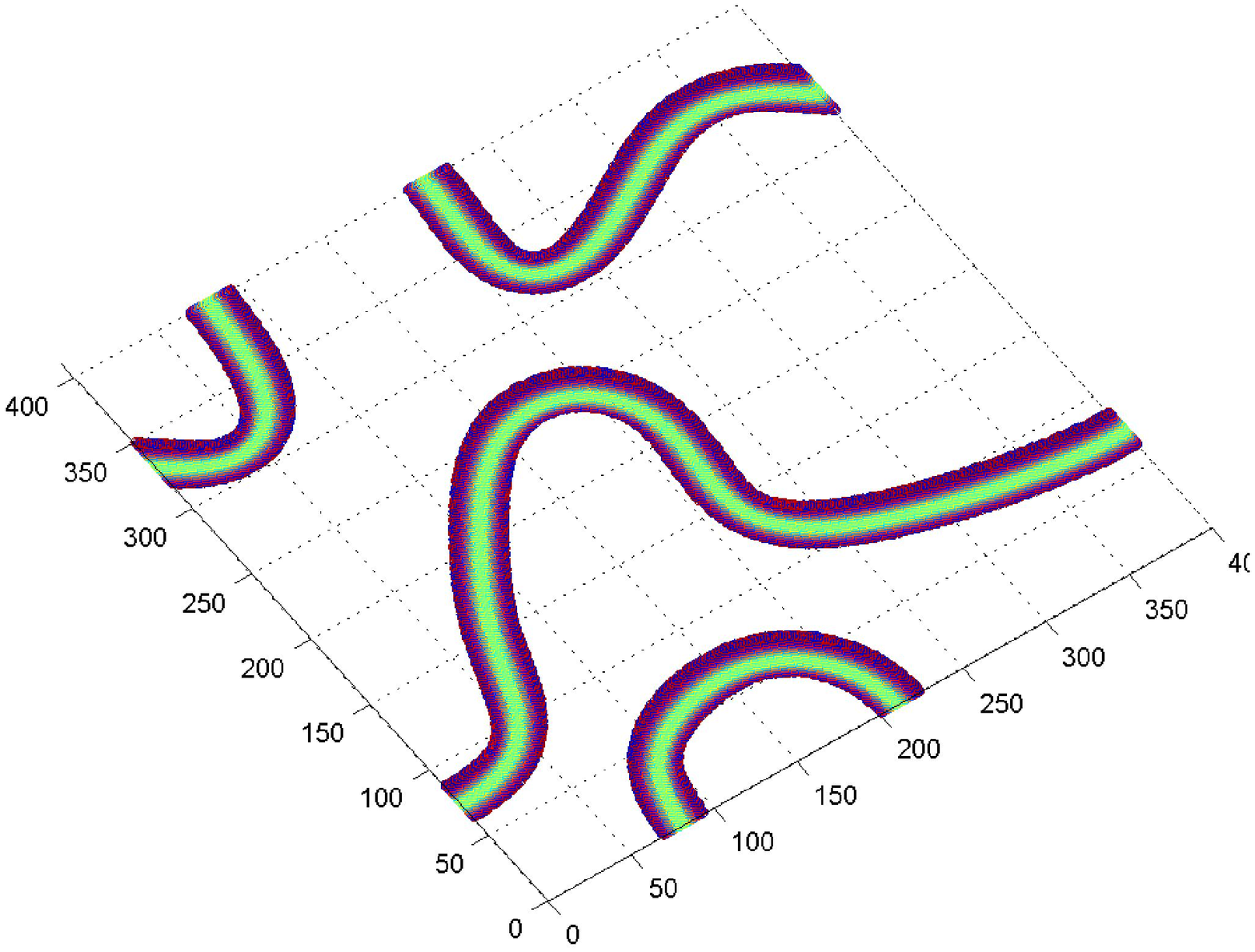}\\
  \caption{Spatial domains in two dimensional system, for logistic growth with nearest neighbors
  competition. The parameters are chosen to be above the bifurcation threshold, and the stable steady
  state is a "checkerboard" with alternating filled and empty sites. Denoting a site by its coordinates $i,j$, there are
  two possible phases of the solution, correspond to filled $i+j \ \ odd$, empty $i+j \ \ even$ and vice versa.
  Here, the results of an Euler integration of the process for a $2d$ sample of $50x50$ sites with
  periodic boundary conditions  is presented, where only the kinks separating regions of different order parameters are
  colored. The kinks here are non-contractible on the torus and correspond to one dimensional topological defects.
  The simulation parameters are $D=0, \ \gamma_1 = 0.2505$. Initial conditions are "seed" population at each site taken
  randomly from a square distribution between $[0,0.01]$.      }\label{kink1d}
\end{figure}

\begin{figure}
  \includegraphics[width=7cm]{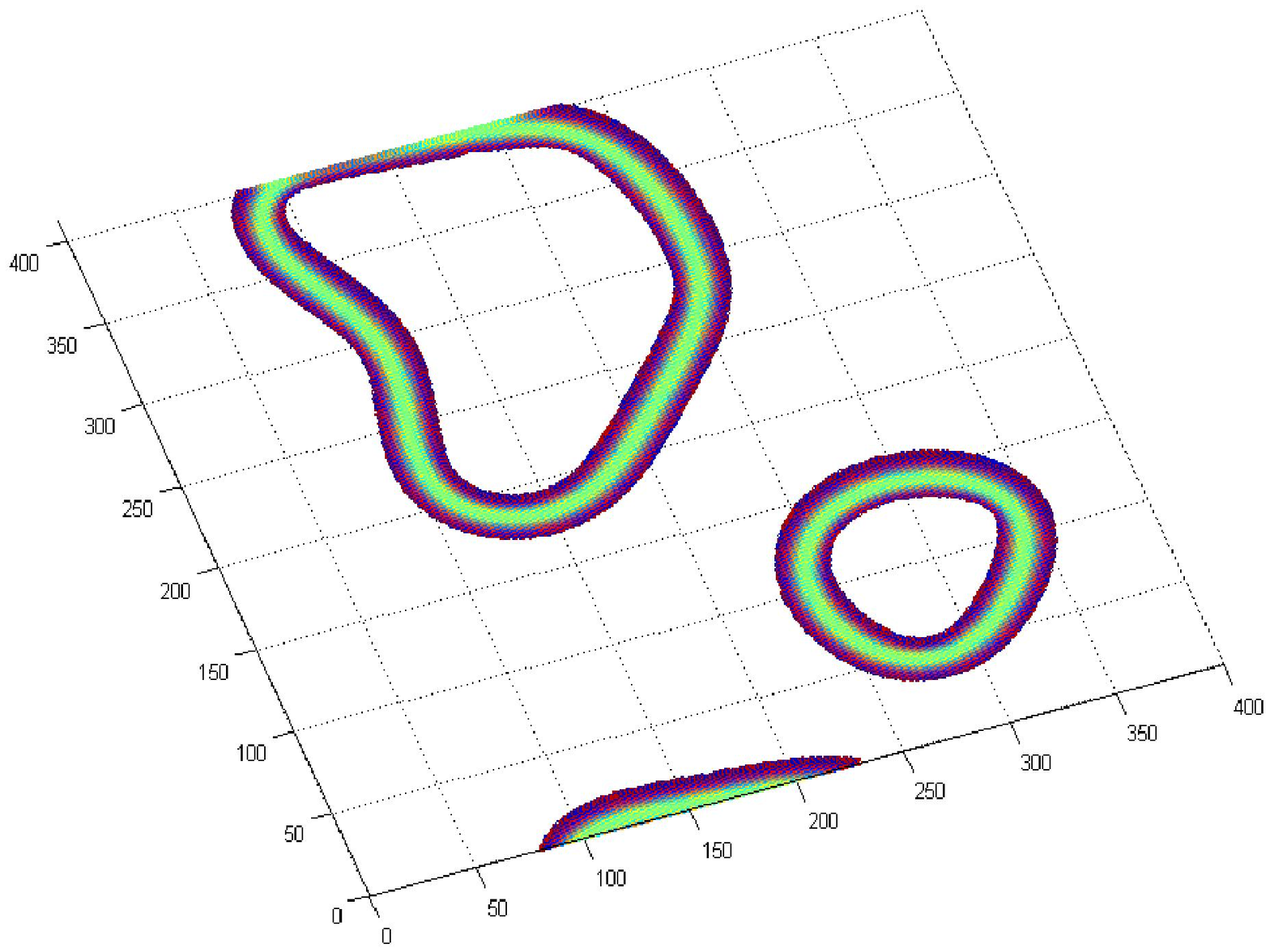}\\
  \caption{The same system and parameters as in Figure (\ref{kink1d}), for other choice of random initial
  conditions. Here the domain wall
  is contractible on the torus and the order parameter phases are different
  between the inside and the outside of the kink.  }\label{kink2d}
\end{figure}

For nearest neighbors interaction kinks between different regions
(checkerboard parity) occurs. Because of the two dimensionality of
the lattice the kinks might have any arbitrary spatial line,
rather then straight  line, as shown at figure (\ref{kink1d}).
Those kinks are de-facto one dimensional topological defects,
because of the periodic boundary conditions. On the other hand,
the domain walls of Figure (\ref{kink2d}) seems to admit a real 2d
features, although their topological character is not clear.

\section{Global properties}

\subsection{Upper critical diffusion}

Let us turn back to the bifurcation condition, Eq. (\ref{bif}), in
different representation:
\begin{equation}\label{bif1}
\frac{\beta_k}{\beta_0} + 2  D [1-cos(kl_0)]< 0
\end{equation}
where the $k$ considered is the one for which $\beta_k$ admits a
global minimum. Clearly, this $k_{min}$ depends only on the form
of the interaction kernel and is independent of its strength (if
one multiplies all $\gamma_r$ by constant factor, the value of
$k_{min}$ remains the same). Since the negative term in the
instability condition $\frac{\beta_k}{\beta_0}$ cannot exceed
$(-1)$, the absolute value of the right hand term should be even
smaller to allow a periodic modulation of the stable steady state.

Assume, now, that the wavelength of the modulation is much larger
than the lattice constant (as already required as one approaches
the  the continuum limit). In that case the approximation $2D
[1-cos(kl_0)] \approx  D k_{min}^2 l_0^2$ holds, and since $D =
(W_F/l_0)^2$, this term is proportional to $(W_F/\lambda)^2$,
where $\lambda$ is the period of the modulation. This implies
that, independent of the strength of the long-range competition,
\emph{bifurcation never takes place if the width of the Fisher
front is larger than the period of the modulation}. This statement
holds up to a numerical factor (between zero and one) which
determined by the form of the competition kernel.

Simple example that demonstrate these considerations is the case
of nearest neighbors interaction. Here
\begin{equation}
g(k)=1+2\gamma cos(k)+2(1+2\gamma)D[1-cos(kl_0)]
\end{equation}
and the  global minima  is  $k=\pi$.  $g(k_{min})$ is
\begin{equation}
g(\pi)=1-2\gamma+4(1+2\gamma)D
\end{equation}
so for any $\gamma$ there is an upper critical D
\begin{equation}
D_c=\frac{2\gamma-1}{4(1+2\gamma)}
\end{equation}
above which no bifurcation takes place. This upper critical
diffusion constant converges to a global value as $\gamma \to
\infty$
 \begin{equation}
D_c^g \equiv {D_c,}_{\gamma->\infty} = \frac{1}{4}.
\end{equation}
and no bifurcation takes place when the width of the Fisher front
is of order of the modulation length. Intuitively this result may
be understood as follows: suppose that the system is in its 010101
state, and suppose that the dynamics is discrete in time. If
$D=1/4$ it implies that each filled site contribute $1/4$ to any
of its neighbors, and then the system is frozen in its homogenous
state with amplitude $1/2$ at each site. Generalizing this
intuition to periodic modulation of arbitrary  wavelength yields
the same result, where the Fisher front width stands as a
definition of an "effective site".

\subsection{Spatial segregation and total population}

Given a  system with long range competition, one may ask how the
\emph{total} population (integrated over all the spatial domain)
or the average population density, depend on the phase of the
system. Naively one expects the total population to grow with the
diffusion constant, as faster spatial wandering helps an
individual reactant to escape from the depleted region of an
already existing colony. This, however, is not the case, as
pointed out by \cite{Birch} and \cite{Garcia}: the size of the
total population depends on the efficiency of segregation: strong
segregation implies higher population (on average, since there are
empty regions and living patches). Thus, decrease of diffusion
implies higher total population density.

Clearly, the total population is given by the amplitude of the
zero mode in Fourier decomposition of the population, (See Eq.
\ref{eq3}). As long as the system is in its homogenous phase this
quantity is diffusion independent and the total population depends
only on the strength of the interaction, $A_0=1/\beta_0$. Right
above the bifurcation, when only one excited mode ($k$) exists,
the total population is proportional to $A_0 = \alpha_k/(\beta_0 +
\beta_k)$, and since $\alpha_k$ increases as $D$ decreases, so is
the total population. In the case of one dimensional lattice with
nearest neighbors interaction, for example, the dependence of the
total occupancy of the sample on the diffusion constant may be
calculated explicitly, since there is only one excited mode
$k=\pi/l_0$. Here even far from the bifurcation point the
amplitude  of the zero mode is given by $A_0 = \alpha_k/(\beta_0 +
\beta_k)$. The total sum versus diffusion is, accordingly,
\begin{equation}
A_0 = \left\{ \begin{array}{cc}
   (1-4D)/2 & D < D_c \\
   1/(1+2 \gamma) & D > D_c.
\end{array} \right.
\end{equation}

\begin{figure}
  \includegraphics[width=7cm]{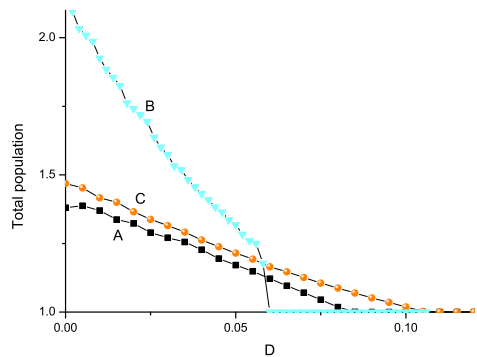}\\
  \caption{Total population versus diffusion coefficient for several situations. (A):  $1d$ with nearest
  neighbors interaction (squares). (C): $1d$, n.n.n. interaction (circles). (B): $2d$,
  top hat interaction (triangles). The "top hat" is constant interaction with all sites inside a circle of radius $3
l_0$, and zero outside. In order to present all the results in the
same panel, the population has been normalized, for each system,
by its homogenous solution. }\label{sum}
\end{figure}

Figure (\ref{sum})  shows the total sum versus diffusion  for few
situations. The numerical results indicate that the decay of
average population is approximately linear. Note that, for the
"top hat" competition presented here, there seems to be a
discontinuity at $D_c$ in two dimensions, while in $1d$ the total
population is continuous at the transition.

\section{Local initiation: the dynamics of invasion and segregation}

Along this paper, an analysis of the stable steady states of the
logistic growth with long range competition was presented. As few
stable steady solutions may exist simultaneity for the same set of
parameters, the generic situation was identified numerically using
global initiation, i.e., small random population at each site. In
this section, the dynamics of growth is analyzed, where the
initial conditions are a colony with compact support. For local
logistic growth this problem was considered years ago by   Fisher
\cite{fisher} and Kolomogorov \cite{kol}. The invasion of the
stable solution into the unstable one takes place via a front (the
Fisher front) that propagates in constant velocity.  This problem
was considered by many authors in different contexts and was
generalized to other cases of invasion into unstable state, see
comprehensive review   by van Saarloos  \cite{saarloos}.

\begin{figure}
  \includegraphics[width=7cm]{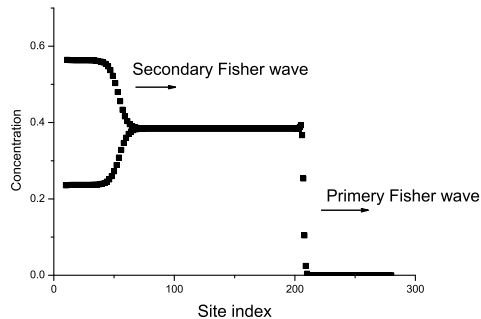}\\
  \caption{Snapshot of the one dimensional system, initiated locally from the left, where the primary
  velocity is higher then the secondary velocity. The two fronts are clearly shown, and the homogenous
  region between them is widening as time elapsed.  The simulation assumes nearest neighbors competition with:
  D=0.04 and $\gamma=0.8$. Along time, defects (not shown) are
  generated at the tip of the secondary front due to the noise
  induced by the primary front.
   }\label{vp}
\end{figure}

As emphasized above, the system considered here may admit [in
region II and III of figure (\ref{fig3})]  two instabilities: the
empty state is unstable against the homogenous one, while the
homogenous solution breaks and yields a spatial modulation.
Accordingly, if the system is initiated locally from a small
colony of compact support one expects that two fronts propagate
into the empty region: first the front associated with the
homogenous state, and then the modulation (secondary instability)
front \cite{hohenberg}.  These two fronts travel in different
velocities. Generally, it is known that  the  Fisher velocity is
determined by the leading edge ("pulled" fronts) and is  related
to the Lyapunov exponent that characterizes the relevant
instability. Accordingly, the dynamics of our system is determined
by two velocities:   $v_p$, the velocity of the primary front
(that interpolates between the empty and the homogenous state) and
the modulation velocity $v_s$. While  $v_p$ is $\gamma$
independent, the secondary front velocity $v_s$ depends on the
characteristics of the long range competition. By tuning of
$\gamma$, though, one may change the relative velocity between the
primary and the secondary front. Both velocities may be calculated
analytically using a saddle point method and taking into account
the discreteness of the lattice points, as discussed in Appendix
A. Generically,  there are two possible scenarios for the takeover
of an empty region by spatially modulated steady state: in the
first case $v_p
> v_s$ (see Figure \ref{vp}) and the homogenous region between the
primary and the secondary front grows linearly in time. This
situation is very sensitive, as small perturbations (induced by
the leading front) lead to spontaneous bifurcation of the
homogenous region, a process that yields many structural defects
(e.g., kinks) along the chain.

\begin{figure}
  \includegraphics[width=7cm]{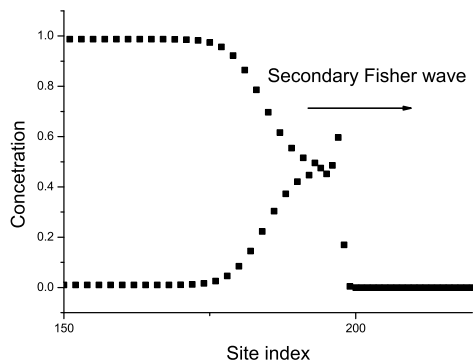}\\
  \caption{Same as Figure \ref{vp}, but now the velocity of the secondary front is
  higher than the velocity of the primary front. Since the secondary instability may appear only
  after the primary, the velocity of the whole front is determined by $v_p$.
  The parameters used  are: D=0.04, $ \gamma=2 $. } \label{vs}
\end{figure}

In the second case the situation is different: if $v_p < v_s$
there is no homogenous region, and only one front exists. Its
velocity is determined, of course, by the primary front velocity,
but its shape is different (see Figure \ref{vs}).  In that case
the sensitive homogenous region never exists, and the pattern
formation process is robust, with no defects associated with the
front kinetics.

\section{Conclusions and remarks}

This paper attempt to present the various phases associated with
the steady states of the logistic process on spatial domains with
nonlocal competition. The main feature is, of course, the
segregation transition that happens, as was shown, where the width
of the Fisher front (associated with the homogenous solution)
becomes shorter than the instability wavelength. Right above the
bifurcation one finds a pattern dominated by a single wavelength,
while far away from the bifurcation line the stable steady state
becomes spiky. Each phase is associated with its own defects:
phase shift close to the bifurcation, empty regions in the spiky
phase, and domain walls (kinks) for the up-down phase of the
nearest-neighbors interaction. It turns out that the segregation
transition increase the overall carrying capacity per unit volume.
In $1d$ the population is continues at the transition while in two
dimensions discontinuity might occur.

Upon local initiation the system dynamics is governed by the
relations between the velocities of the primary (empty to
homogenous) and the secondary (homogenous to modulated) fronts.
The numerics suggests that, while global initiation may yield
"disordered" structure with many defects per unit length, local
initiation with the same parameters yields ordered structure
unless the secondary front velocity is smaller that the primary
one.

While in this work only rate equations of reaction-diffusion type
has been considered, in recent numerical works of Birch and Young
\cite{Birch} and Garcia et. al. \cite{Garcia}  the stochastic
motion of the individual reactants is taken  into account. These
stochastic models add two ingredients to the description presented
here. First, the introduction  of individual reactants ("Brownian
bugs" \cite{young2}) implies a \emph{threshold} on the reactant
concentration on single patch. Secund, there is a multiplicative
noise associated with the stochastic motion of individual
reactants. As shown in this work, many of the features associated
with long range competition are independent of the discrete nature
of individual reactants.

\section{acknowledgements}
The authors thank Prof. David Kessler for many helpful
discussions. This work was supported by the Israeli Science
Foundation,  grant no. 281/03, and by Yeshaya Horowitz Fellowship.

\section{Appendix A}

In this appendix the analytic expression for the secondary front
velocity on a discrete lattice is obtained, via the saddle point
argument (see  \cite{levin}). For the sake of simplicity, only the
case of nearest neighbors interaction is considered. In order to
preform the same calculations for competition beyond the n.n.
limit, one should first find numerically the steady state
modulation and then follow the same procedure.

The evolution of a population is given by:
\begin{equation}\label{21}
\frac{\partial c_n}{\partial t}= D [-2c_n+c_{n+1}+c_{n-1}] + c_n -
c_n^2 + c_n\gamma(c_{n+1}+c_{n-1}).
\end{equation}
Denoting by $\delta_n$ the deviations from the homogenous
solution,  $c_n=A_0+\delta_n$, Eq. (\ref{21}) is linearized to
yield:
\begin{equation}\label{22}
\frac{\partial \delta_n}{\partial t}= \alpha \delta_n +\beta
(\delta_{n+1}+\delta_{n-1})
\end{equation}
Where $\alpha= a-2bA_0+2A_0\gamma-2D/l_0^2$ and $
\beta=D/l_0^2-A_0\gamma$. Assuming  modulated solution  of the
form,
\begin{equation} \label{23}
\delta_n = \left\{ \begin{array}{cc}
  Ae^{ikl_0n+\Gamma(k)t} & \ \  n \  odd   \\
  Be^{ikl_0n+\Gamma(k)t} & \ \  n  \  even  \\
\end{array} \right.
\end{equation}
and plugging (\ref{23}) into (\ref{22}) one gets
\begin{equation}
\Gamma(k) \left[ \begin{array}{c}
  A \\
  B \\
\end{array} \right] =
\left[ \begin{array}{cc}
  \alpha & \beta \cos(kl_o) \\
  \beta \cos(kl_o) & \alpha \\
\end{array}\right]\left[ \begin{array}{c}
  A \\
  B \\
\end{array} \right].
\end{equation}
The dispersion relations is given by:
\begin{equation}
\Gamma(k)=\alpha + \beta \cos(kl_0).
\end{equation}
where the plus sign is chosen for the unstable modes. The
solutions are  of the form
\begin{equation}
\left[ \begin{array}{c}
 c_n \\
  c_{n+1}  \\
\end{array} \right] =\left[  \begin{array}{c}
 A \\
  B \\
\end{array} \right] e^{ikx+\Gamma(k)t}.
\end{equation}

\begin{figure}
  \includegraphics[width=7cm]{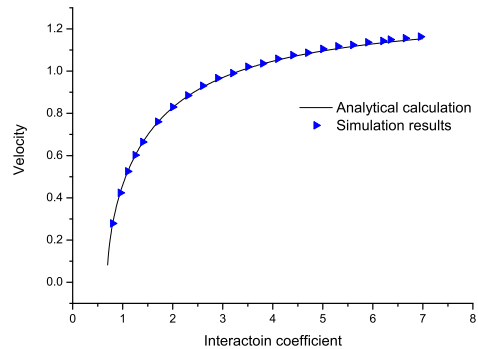}\\
  \caption{Comparison of the numerical simulation (triangles) and the
  theoretical prediction based on the saddle point method [Eqs. (\ref{33}) and (\ref{34}), solid line]
  for the velocity  of the secondary front as a function of the interaction strength. The diffusion used is
    $D=0.04$ and the lattice constant is $l_0 =1$, $dt = 0.01$  }\label{vp2}
\end{figure}

If a solution represents a travelling front with velocity $v$ it
is useful to define the coordinate system in the moving frame,
$\zeta=x-vt$, to get
\begin{equation}
\left[ \begin{array}{c}
 c_n \\
  c_{n+1}  \\
\end{array} \right] = \left[ \begin{array}{c}
  A \\
  B \\
\end{array}\right] e^{ik\zeta+ikvt+\Gamma(k)t}.
\end{equation}
Using the saddle point method \cite{saarloos} the two equations
that determine the velocity are
\begin{equation}\label{33}
f \equiv ivk+\alpha+2\beta cosh(kl_0)=0
\end{equation}
and
\begin{equation}\label{34}
\frac{\partial f}{\partial k} = iv+ 2 \beta_0 sinh(kl_0) = 0.
\end{equation}
In case of finite time steps one should replace  $ivk$ by $
(e^{-ikv dt}-1)/ dt$  to get the appropriate corrections. Figure
(\ref{vp2}) shows the perfect fit between the solution of
(\ref{33}) and (\ref{34}) and the numerical solution.

\section{Acknowledgements}

We thank Prof. David Kessler for many helpful discussions and
comments. This work was supported by the Israel Science Foundation
(grant no. 281/03) and by the Yeshaya Horowitz Fellowship.

\end{document}